\documentclass[11pt,a4paper]{article}
\usepackage[utf8]{inputenc}
\usepackage{geometry}
\usepackage{amsmath}
\usepackage{amsfonts}
\usepackage{amssymb}
\usepackage{graphicx}
\usepackage{cite}
\usepackage{url}
\usepackage{hyperref}
\usepackage{multirow}
\usepackage{adjustbox}
\usepackage{appendix}
\usepackage{subcaption}
\usepackage{authblk} 

\title{SALPA: Spaceborne LiDAR Point Adjustment for Enhanced GEDI Footprint Geolocation}
\author[1]{Narumasa Tsutsumida\thanks{Corresponding author. Email: rsnaru.jp@gmail.com}}
\author[2]{Rei Mitsuhashi}
\author[2]{Yoshito Sawada}
\author[3]{Akira Kato}

\affil[1]{Graduate School of Science \& Engineering, Saitama University, Japan}
\affil[2]{Japan Aerospace Exploration Agency, Japan}
\affil[3]{Graduate School of Horticulture, Chiba University, Japan }

\date{\today}

\begin{document}

\maketitle
\begin{abstract}
Spaceborne Light Detection and Ranging (LiDAR) systems, such as NASA's Global Ecosystem Dynamics Investigation (GEDI), provide 
forest structure for global carbon assessments. However, geolocation uncertainties (typically 5--15~m) propagate systematically through derived products, undermining forest profile estimates, including carbon stock assessments. Existing 
correction methods face critical limitations: waveform simulation approaches achieve meter-level accuracy but require high-resolution LiDAR data unavailable in most regions, while terrain-based methods employ deterministic grid searches that may overlook optimal solutions in continuous solution spaces. We present SALPA (Spaceborne LiDAR Point Adjustment), a multi-algorithm optimization framework integrating three optimization paradigms with five distance metrics. Operating exclusively with globally available digital elevation models and geoid data, SALPA explores continuous solution spaces through gradient-based, evolutionary, and swarm intelligence approaches. Validation across contrasting 
sites: topographically complex Nikko, Japan, and flat Landes, France, demonstrates 15--16\% improvements over original GEDI positions and 0.5--2\% improvements over the state-of-the-art GeoGEDI algorithm. L-BFGS-B with Area-based metrics achieves 
optimal accuracy-efficiency trade-offs, while population-based algorithms (genetic algorithms, particle swarm optimization) excel in complex terrain. The platform-agnostic framework facilitates straightforward adaptation to emerging spaceborne LiDAR missions, providing a generalizable foundation for universal geolocation correction essential for 
reliable global forest monitoring and climate policy decisions.
\end{abstract}

\section{Introduction}

Spaceborne Light Detection and Ranging (LiDAR) systems have fundamentally transformed Earth observation by enabling direct three-dimensional measurements of terrestrial ecosystems across global scales \cite{Simard2011-ud,Dubayah2020-py,Abshire2005-hd,de-Conto2024-pb,Magruder2024-ob}. Unlike passive optical sensors, LiDAR systems actively emit laser pulses or measure photon time of flight~\cite{Dubayah2020-py,de-Conto2024-pb,Ceccherini2023-uv,Neuenschwander2019-fa,Markus2017-nn}. These technologies deliver critical measurements of vertical forest structure \cite{Burns2024-un,Ceccherini2023-uv,Lang2022-ep,Moudry2024-cw,Wang2024-cb,de-Conto2024-pb}, surface elevation \cite{Mitsuhashi2024-po,Moudry2024-cw,Adam2020-kk,Li2024-qw}, and canopy architecture \cite{Potapov2021-jr,Schneider2020-ho} across space. They quantify climate variables \cite{Magruder2024-ob}, terrestrial carbon stocks \cite{Nazir2025-mu, Kellner2023-hm,Cushman2023-ey, Sun2022-ch, Chen2022-sg, Hernandez-Martinez2025-kx, Duncanson2022-wt,Li2024-qw}, and biodiversity \cite{Torresani2023-nv,Marselis2019-kt} that are unattainable through conventional remote sensing and limited field observations.

The Global Ecosystem Dynamics Investigation (GEDI) exemplifies both the potential and challenges of International Space Station (ISS) based LiDAR systems. Operating as a full waveform instrument, GEDI provides forest structure measurements between 51.6°N and 51.6°S, covering over 85\% of Earth's forests~\cite{Dubayah2020-py}. Since April 2019, GEDI has acquired over 15 billion observations with 25-meter footprints, enabling wall-to-wall forest mapping and continental-scale carbon assessments~\cite{Kellner2023-hm, Potapov2021-jr}. However, ISS mounting introduces fundamental geolocation challenges: lacking dedicated orbit determination systems, these instruments inherit positioning uncertainties from spacecraft attitude variations, structural vibrations, and thermal deformations~\cite{Tang2023-ku, Yang2024-tn}. As space agencies expand capabilities, including JAXA's planned Multi-sensing Observation Lidar and Imager (MOLI) \cite{Sakae2025-hl}, robust post-processing correction methodologies applicable across diverse spaceborne LiDAR architectures are critically needed.

Geolocation uncertainties fundamentally constrain spaceborne LiDAR utility across all platforms. For GEDI, horizontal errors typically range from 5 to 15 meters~\cite{Potapov2021-jr, Schleich2023-ax}, propagating systematically through derived products and undermining carbon assessments~\cite{Chen2025-wf, Li2024-lj}. In heterogeneous landscapes, small displacements (5 to 10 m) fundamentally alter sampled characteristics, potentially misclassifying forest types or introducing false change signals~\cite{Schleich2023-ax, Tsutsumida2025-se}. Even dedicated satellite missions require post-processing corrections in complex terrain~\cite{Luthcke2021-nf}.

Existing correction methodologies face critical limitations. Waveform simulation approaches achieve centimeter-level accuracy but require high-resolution LiDAR data, restricting applicability~\cite{Xu2023-yv, Mitsuhashi2024-po}. Machine learning methods show improvements but exhibit limited generalizability and require extensive training data~\cite{Mitsuhashi2024-po}. Terrain-based optimization (GeoGEDI~\cite{Schleich2023-ax}) offers broader applicability through digital elevation model (DEM)-based elevation matching, achieving  geolocational error reductions. However, the current deterministic grid searches may overlook optimal solutions and rely on single objective metrics without exploiting terrain morphology or spatial patterns~\cite{Tang2023-ku}.

This study presents SALPA (Spaceborne LiDAR Point Adjustment), a multi-algorithm optimization framework for universal spaceborne LiDAR geolocation correction. SALPA integrates three optimization paradigms (L-BFGS-B, Genetic Algorithms (GA), and Particle Swarm Optimization (PSO)) with five distance metrics (Euclidean, Manhattan, Hausdorff, area-based, and correlation-based). Operating exclusively with DEMs and geoid models, SALPA ensures global deployability while exploring continuous solution spaces inaccessible to grid search methods. The modular architecture facilitates straightforward adaptation to emerging platforms through simple parameterization.

In this study, we pursue two objectives: (1) develop and validate a platform-agnostic framework achieving competitive accuracy using globally available data, demonstrated through GEDI validation; and (2) rigorously compare optimization algorithms across contrasting terrain conditions. Validation encompasses two contrasting sites: topographically complex Nikko, Japan, and flat Landes, France.

\section{Related Work}

Two approaches are widely used for spaceborne LiDAR geolocation correction: waveform simulation and terrain-based optimization.

\subsection{Waveform Simulation Methods}

Waveform-based correction matches observed LiDAR returns with synthetic waveforms generated from airborne or ground-based LiDAR point clouds at candidate positions around nominal footprint locations~\cite{Xu2023-yv, Hancock2019-tb}. For each footprint, the position that maximizes waveform similarity (typically via correlation coefficients) becomes the corrected location. Using the GEDI simulator framework~\cite{Hancock2019-tb} with Simplex optimization, Xu et al.~\cite{Xu2023-yv} achieved high accuracy when similarity coefficients exceeded 0.80. Their validation showed that 32\% of weak-orbit GEDI datasets could be upgraded to good quality through optimized thresholds, reducing mean horizontal errors to 9.46 m for good orbit data.

Machine learning offers an alternative to physical simulation. Mitsuhashi et al.~\cite{Mitsuhashi2024-po} developed deep learning models that estimate ground elevation directly from waveform characteristics, implicitly correcting geolocation errors in ground detection. Their approach achieved $>$3~m RMSE improvements over operational GEDI L2A products across subarctic to tropical regions. However, these data-driven methods face generalization challenges when applied to ecosystems that differ from training domains in species composition or structure, requiring extensive labeled datasets unavailable in most regions~\cite{Li2024-lj}.

Both physical simulation and machine learning waveform approaches require high-quality LiDAR point clouds that capture complete three-dimensional forest structure. This restricts applicability to well-surveyed regions, predominantly in developed countries~\cite{Xu2023-yv, Mitsuhashi2024-po}, limiting deployment for global monitoring where observations are most needed—in data-sparse tropical and subtropical regions. Additionally, the computational demands of simulating waveforms from LiDAR point clouds constrain scalability for continental applications.

\subsection{Terrain-Based Optimization Methods}

Terrain-based correction offers broader geographic applicability through direct elevation matching between LiDAR observations and reference DEMs, requiring only DEM data rather than detailed vegetation structure. The GeoGEDI algorithm represents the current state-of-the-art, implementing deterministic grid search within ±25~m windows (5~m spacing) that evaluates 121 candidate positions per shot group within a beam~\cite{Schleich2023-ax}. For each offset, GeoGEDI computes the mean absolute error (MAE) between GEDI ground estimates ($elev\_lowestmode$) and DEM values, selecting the position with the minimum discrepancy. Advanced implementations incorporate flow accumulation techniques~\cite{Freeman1991-gb} for topographic feature identification and use cluster-based temporal correlation corrections to address ISS structural vibrations, achieving RMSE reductions up to 59.6\% in mountainous terrain~\cite{Schleich2023-ax}.

Other approaches have explored alternative optimization strategies. Yang et al.~\cite{Yang2024-tn} constructed three-dimensional error distribution matrices, using ensemble averaging across multiple centroid estimation methods (minimum error, center of gravity, 2D Gaussian fitting) while decomposing errors into along-track and cross-track components for systematic bias characterization. Schleich et al.~\cite{Schleich2023-ax} showed that single-beam corrections outperform four-beam approaches due to reduced correlation effects. Validation against airborne LiDAR demonstrated improvements from 0.34–4.23~m for version 1 data and 1.41–3.82~m RMSE for GEDI version 2 data, depending on terrain complexity. Statistical frameworks have quantified geolocation error spatial patterns, revealing systematic biases that vary along and across orbital tracks at beam and orbit scales~\cite{Yang2024-tn}.

While machine learning shows promise for other LiDAR applications, such as canopy height estimation through integration with multispectral imagery~\cite{Wang2024-cb} and uncertainty quantification~\cite{Lang2022-ep}, its application to terrain-based geolocation correction remains limited. The primary advantage of terrain-based methods, namely universal deployability with available DEMs, would be compromised by machine learning's requirement for extensive training datasets and challenges in cross-platform generalization.

Despite their computational efficiency and operational deployability, deterministic terrain-based approaches evaluate only discretized position candidates, potentially overlooking globally optimal solutions in continuous solution spaces. Furthermore, exclusive reliance on elevation discrepancy ignores terrain morphology, spatial patterns, and alternative similarity measures that may capture complementary aspects of terrain-LiDAR correspondence~\cite{Tang2023-ku}. The fixed grid-search strategy, while exhaustive within sampled positions, cannot exploit gradient information or adaptive search strategies that may accelerate convergence in smooth error surfaces.

\subsection{Research Gap and Motivation}

Current geolocation correction methodologies exhibit a fundamental trade-off between accuracy and deployability. Waveform simulation approaches achieve meter-level accuracy but require high-density LiDAR point clouds, restricting applicability to well-surveyed regions~\cite{Xu2023-yv, Mitsuhashi2024-po}. Machine learning extensions demonstrate cross-biome transferability but demand extensive training datasets and exhibit limited platform generalization~\cite{Mitsuhashi2024-po, Li2024-lj}. Conversely, terrain-based methods offer universal deployability through globally available DEMs but employ deterministic grid-search strategies that sample only discretized positions, potentially overlooking optimal solutions in continuous solution spaces~\cite{Schleich2023-ax, Yang2024-tn}. These fixed-grid approaches cannot exploit gradient information in smooth error surfaces or implement adaptive search strategies that may accelerate convergence in complex topography.

Previous studies have confined terrain-based approaches to minimizing a single distance metric (typically Euclidean distance), ignoring potential advantages of alternative similarity measures that capture complementary aspects of terrain-LiDAR correspondence. Furthermore, grid-search strategies for spaceborne LiDAR geolocation correction can be replaced with advanced optimization paradigms~\cite{Gad2022-ul, Shami2022-ke}. The investigation of multiple optimization strategies with diverse distance metrics remains unexplored, despite theoretical expectations that different algorithms exhibit complementary performance characteristics across varying problem landscapes~\cite{Wolpert1997-tu}.

Two critical questions emerge: (1) Can gradient-based, evolutionary, and swarm intelligence optimization paradigms achieve superior accuracy compared to exhaustive grid-search while maintaining computational feasibility? (2) Does the selection of distance metrics (Manhattan, Hausdorff, correlation-based, area-based) influence correction performance across varying terrain complexity?
%(3) How does algorithm performance degrade with decreasing DEM spatial resolution, and can sophisticated optimization compensate for coarser globally available products (30 m) to achieve accuracy competitive with high-resolution local DEMs?

SALPA addresses these gaps by applying multi-algorithm optimization with multiple distance metrics, requiring only available DEMs and geoid models as reference, while exploring continuous solution spaces inaccessible to conventional grid-search. The platform-agnostic framework enables straightforward adaptation to emerging spaceborne LiDAR missions through parameterization adjustments without algorithmic redesign, providing a generalizable foundation for universal geolocation correction.

\section{Methods}

\subsection{Overview}

\begin{figure}[!ht]
    \centering
    \includegraphics[width=1\textwidth]{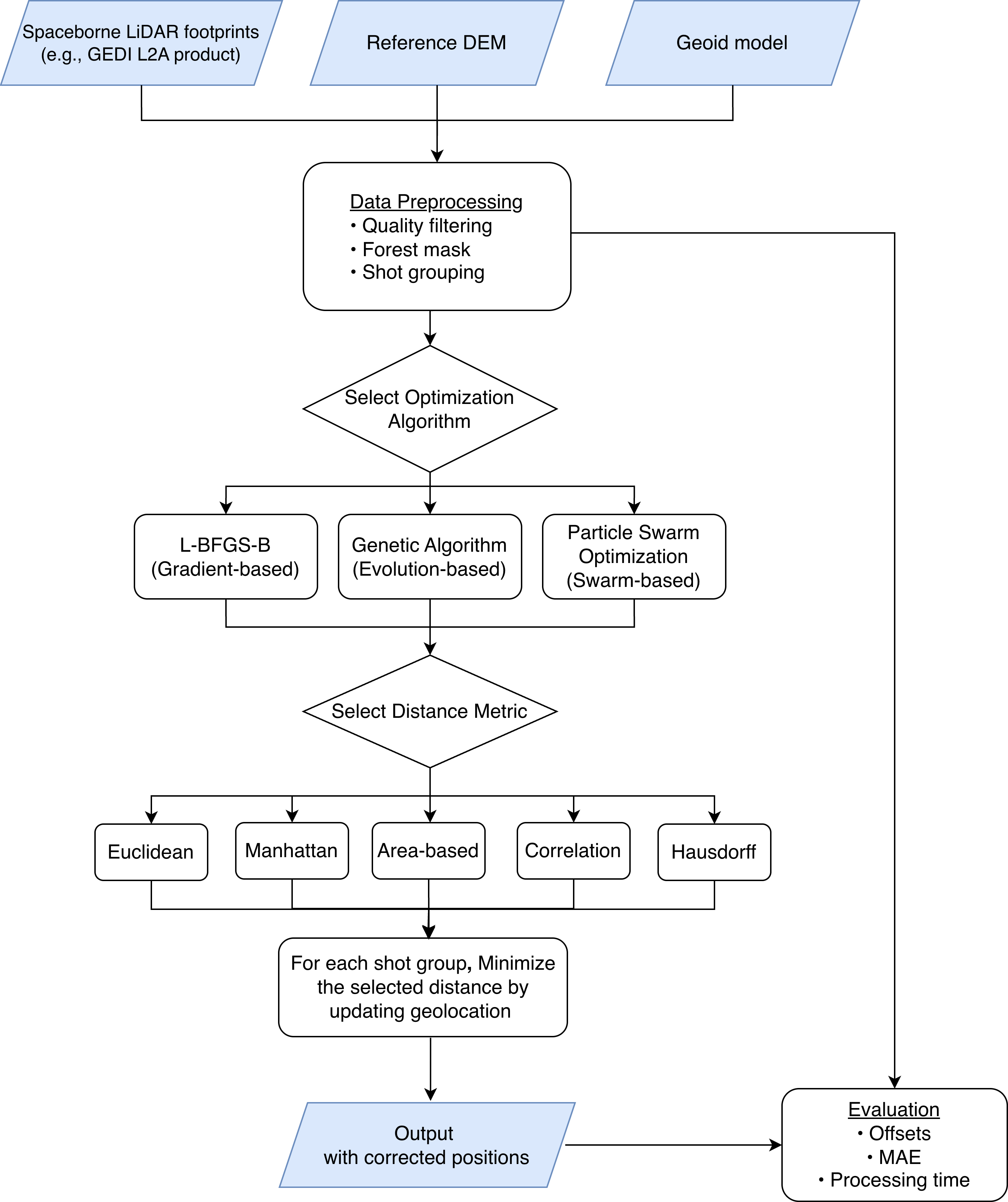}
    \caption{Overview of the SALPA algorithm.}
    \label{fig:salpa_overview}
\end{figure}

SALPA is a novel algorithm that enhances GEDI footprint geolocation through terrain-optimized positioning using DEMs by minimizing the discrepancy between observed GEDI elevations and reference DEM values through optimal spatial displacement. This is achieved by integrating multiple optimization algorithms with diverse distance metrics to identify optimal footprint positions. We only required referenced DEM and geoid data for the application of the SALPA algorithm. The overview of the SALPA algorithm is shown in Figure~\ref{fig:salpa_overview}.

\subsection{SALPA Algorithm Framework}

SALPA formulates geolocation correction as a continuous two-dimensional optimization problem that minimizes terrain elevation discrepancy through optimal spatial displacement. The framework consists of three integrated components: (1) terrain surface modeling, (2) multi-algorithm optimization engine, and (3) comprehensive distance metric evaluation.

\subsubsection{Rationale}

For a shot group $g$ containing $n$ GEDI footprints, let $\mathbf{E}_g = \{e_1, e_2, \ldots, e_n\}$ represent the LiDAR-derived ground elevations and $\mathbf{P}_g = \{(x_1, y_1), (x_2, y_2), \ldots, (x_n, y_n)\}$ represent their corresponding geographic coordinates. The optimization objective seeks the optimal displacement vector $\boldsymbol{\delta} = (\delta_x, \delta_y)$ that minimizes the discrepancy between observed and reference elevations:

\begin{equation}
\boldsymbol{\delta}^* = \arg\min_{\boldsymbol{\delta}} D(\mathbf{E}_g, \mathbf{R}_g(\mathbf{P}_g + \boldsymbol{\delta}))
\label{eq:optimization_objective}
\end{equation}

where $\mathbf{R}_g(\mathbf{P}_g + \boldsymbol{\delta})$ represents reference DEM elevations at displaced positions and $D(\cdot, \cdot)$ denotes the distance metric function. The displacement is constrained within a realistic search window: $|\delta_x|, |\delta_y| \leq 25$ meters, based on expected GEDI positioning uncertainties.

\subsubsection{Reference Terrain Modeling}

For each footprint $i$ at position $(x_i + \delta_x, y_i + \delta_y)$, we extract the corresponding reference elevation using spatial aggregation within a circular buffer of radius $r$. The reference elevation is computed as:

\begin{equation}
r_i(\boldsymbol{\delta}) = \text{agg}\{z(u,v) : \|(u,v) - (x_i + \delta_x, y_i + \delta_y)\| \leq r\}
\label{eq:reference_elevation}
\end{equation}

where $z(u,v)$ represents the DEM elevation at coordinates $(u,v)$ and $\text{agg}(\cdot)$ denotes the aggregation function (mean, median, or mode). While this spatial aggregation accounts for the finite footprint size and provides robust elevation estimates in the presence of DEM noise or resolution limitations, we report the mean aggregation result in this study.

\subsubsection{Distance Metrics}

SALPA implements five complementary distance metrics to capture different aspects of terrain-LiDAR correspondence:

\begin{align}
D_{\text{Euclidean}}(\mathbf{E}, \mathbf{R}) &= \sqrt{\sum_{i=1}^{n} (e_i - r_i)^2} \label{eq:euclidean}\\
D_{\text{Manhattan}}(\mathbf{E}, \mathbf{R}) &= \sum_{i=1}^{n} |e_i - r_i| \label{eq:manhattan}\\
D_{\text{Hausdorff}}(\mathbf{E}, \mathbf{R}) &= \max_{i=1}^{n} |e_i - r_i| \label{eq:hausdorff}\\
D_{\text{Area}}(\mathbf{E}, \mathbf{R}) &= \left|\sum_{i=1}^{n} (e_i - r_i)\right| \label{eq:area} \\
D_{\text{Correlation}}(\mathbf{E}, \mathbf{R}) &= 1 - \frac{\sum_{i=1}^{n}(e_i - \bar{e})(r_i - \bar{r})}{\sqrt{\sum_{i=1}^{n}(e_i - \bar{e})^2}\sqrt{\sum_{i=1}^{n}(r_i - \bar{r})^2}} \label{eq:correlation}
\end{align}

where $\bar{e}$ and $\bar{r}$ represent LiDAR-derived elevation and the spatially aggregated referenced elevation, respectively.

\subsubsection{Optimization Algorithms}

SALPA integrates four optimization paradigms with complementary characteristics for diverse terrain conditions to solve the equation (\ref{eq:optimization_objective}):

\textbf{L-BFGS-B (Limited-memory Broyden-Fletcher-Goldfarb-Shanno Bounded):} A quasi-Newton method employing gradient approximation for rapid convergence in smooth objective landscapes. The algorithm maintains bounded constraints $|\boldsymbol{\delta}| \leq 25$ meters while exploiting local gradient information for efficient search~\cite{Zhu1997-jc}. We set parameters: maximum iterations (100), convergence tolerance ($10^{-6}$).

\textbf{Genetic Algorithm (GA):} A population-based evolutionary method maintaining diversity through selection, crossover, and mutation operations. GA excels in multi-modal terrain landscapes by maintaining multiple candidate solutions simultaneously. We set parameters: population size (50), generations (100), crossover rate (0.8), and mutation rate (0.1).

\textbf{Particle Swarm Optimization (PSO):} A swarm intelligence approach balancing exploration and exploitation through velocity-position updates guided by individual and social learning. PSO demonstrates robust performance across varying terrain complexity with moderate computational requirements. Parameters: swarm size (50), iterations (100), cognitive/social coefficients (1.5), inertia weight (0.5).

% \textbf{Whale Optimization Algorithm (WOA):} A bio-inspired metaheuristic alternating between encircling, bubble-net, and random search phases to avoid local optima. WOA provides excellent exploration capabilities in complex optimization landscapes. Parameters: population size (50), iterations (100), exploitation probability (0.5).

The framework outputs optimal displacement vectors for each shot group, enabling systematic correction of positioning errors while maintaining computational tractability for operational deployment. SALPA is implemented as an open-source R package~\cite{tsutsumida2025salpa}, ensuring reproducibility and community adoption.

\section{Experiments}
\subsection{Study Areas}

\begin{figure}[ht]
    \centering
    \includegraphics[width=\textwidth]{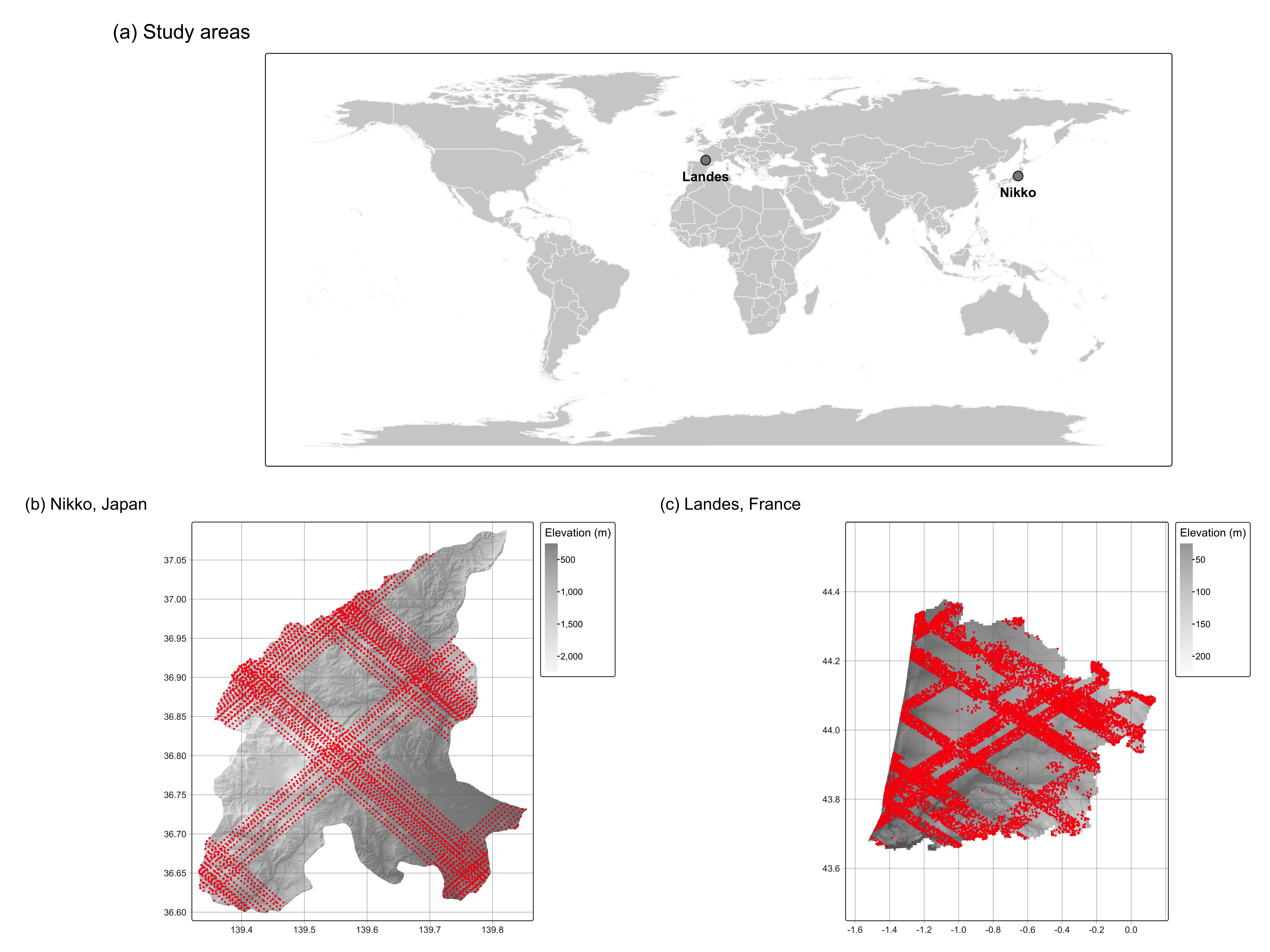}
    \caption{Study areas used for SALPA algorithm evaluation showing contrasting topographic conditions: (a) complex mountainous terrain in Nikko, Japan, and (b) low-relief managed forest landscape in Landes, France.}
    \label{fig:study_areas}
\end{figure}

This study was conducted across two contrasting geographic regions to evaluate the robustness and generalizability of the SALPA algorithm. The Nikko study area, located in the mountainous regions of central Japan (36.5°--37.0°N, 139.0°--139.8°E), is characterized by mixed temperate forests with complex topography featuring steep slopes and elevation variations ranging from 400 to 2,500~m (Figure~\ref{fig:study_areas}a). Nikko was selected due to its complex terrain, which provides an ideal testbed for evaluating geolocation algorithms under challenging topographic conditions.

The Landes study area, situated in southwestern France (44.0°--44.8°N, 1.2°W--0.8°E), represents a contrasting landscape dominated by extensive managed pine forest plantations on relatively flat terrain with minimal elevation variation ($<$200~m) (Figure~\ref{fig:study_areas}b). This popular experimental site was selected to validate SALPA's performance in low-relief environments where traditional elevation-based correction methods may be less effective.

\subsection{Data Sources}
\subsubsection{GEDI L2A Shot Data}
The primary source of shot data was the NASA GEDI Level 2A (L2A) product, which provides detailed elevation and waveform metrics for each laser shot. Key variables extracted included \texttt{elev\_lowestmode}, \texttt{shot\_number}, \texttt{beam}, and \texttt{quality\_flag}.

For the Nikko study area, the GEDI L2A data were collected from April 2021 to October 2021, for a total of 58,736 shots.
For the Landes study area, the GEDI L2A data were collected from May 2019 to May 2020, for a total of 162,856 shots.

\subsubsection{Digital Elevation Model Data}

High-resolution DEMs were used as reference surfaces for positional correction. The Nikko, High-resolution airborne LiDAR-derived DEMs were available at 0.5~m (https://test.geospatial.jp/ckan/dataset/dem05\_tochigi).
For Landes, a 1~m resolution airborne LiDAR DEM was employed for this region, available at https://geoservices.ign.fr/rgealti.
To correct for the effect of the Earth's curvature, region-specific geoid models: GSIGEO2011 for Nikko and RAD09 for Landes were used.

\subsubsection{Land Cover Data}
To mask Tree cover areas only, as GEDI shot data are well observed in forested areas, while each shot in unforested areas, such as urban areas, may be inaccurate for estimating the ground elevation as $elev\_lowestmode$, we used the land cover data from the European Space Agency (ESA) WorldCover product, version 100 \cite{Zanaga2021-xd}. We used the tree cover data (`100`) to mask the GEDI shots in the Nikko study area and the Landes study area.

\subsection{Data Preprocessing}

A series of filters was applied to the downloaded GEDI shots to ensure only high-quality, relevant data were retained for analysis. Specifically, the following criteria were enforced:
\begin{itemize}
    \item \texttt{elev\_lowestmode} $>$ 0 and $<$ 2500~m (to exclude invalid or extreme elevations)
    \item \texttt{degrade\_flag} = 0 (no instrument degradation)
    \item \texttt{quality\_flag} = 1 (highest quality)
    \item \texttt{sensitivity} $\geq$ 0.95 (high waveform sensitivity)
    \item \texttt{rh100} $>$ 0 (positive canopy height)
\end{itemize}

While we used this preprocessed GEDI L2A data, there may be yet outlier data that influences the geolocation adjustment by GeoGEDI and SALPA. 
To further improve data quality, a rolling window outlier detection algorithm was applied to the \texttt{elev\_lowestmode} variable within each trajectory (shot group). For each shot, a rolling mean and standard deviation were computed over a window of 7 shots. Shots deviating by more than 2 standard deviations from the local mean were flagged as outliers and excluded from further analysis.
Additionally, shots were spatially filtered to ensure they fell within the DEM coverage, and any with missing or NA values in key fields were excluded.

Shots were grouped by a unique identifier (typically the first 10 characters of the shot number, referred to as `shot\_group`) to facilitate group-wise adjustment, as required by both GeoGEDI and SALPA. All spatial data were reprojected to the DEM's CRS (e.g., UTM zone 30N for Landes) to ensure spatial consistency in subsequent analyses.

For each valid shot, the geoid undulation was extracted from the reprojected geoid raster and subtracted from the \texttt{elev\_lowestmode} to yield a DEM-referenced elevation (\texttt{gedi\_dem}). The DEM value at each shot location was extracted using a buffer of $r=12.5$, matching the GEDI footprint radius, and the mean DEM value within the buffer was computed (eq. (\ref{eq:reference_elevation}). Shots with a difference between the GEDI-derived elevation and the DEM exceeding 50~m were excluded as likely mismatches or gross errors.

% References:
% \cite{aschleich2023geogedi} A. Schleich et al., "Improving GEDI Footprint Geolocation Using a High-Resolution Digital Elevation Model," IEEE JSTARS, 2023.
% \cite{tsutsumida2025salpa} N. Tsutsumida, "salpa: Satellite LiDAR Point Adjustment," R package, 2025.

\subsection{Evaluation}

The accuracy of each positional adjustment method was assessed using MAE against the referenced DEM (eq.~\ref{eq:mae}).

\begin{equation}
    \label{eq:mae}
    \text{MAE} = \frac{1}{N} \sum_{i=1}^{N} |e_i - r_i|, \\
\end{equation}

where $e_i$ is the GEDI-derived ground elevation for footprint $i$, $r_i$ is the corresponding reference DEM elevation, and $N$ is the total number of footprints in the evaluation dataset.

Spatial displacement was reported to quantify the degree of misalignment per shot group (eq.~\ref{eq:displacement}).

\begin{equation}
\label{eq:displacement}
    \text{Spatial Displacement}_g = \sqrt{(\delta_{x,g})^2 + (\delta_{y,g})^2}
\end{equation}

where $\delta_{x,g}$ and $\delta_{y,g}$ are the optimal displacement values (in meters) determined by the optimization algorithm for shot group $g$.

\subsection{Comparative Framework}
To validate the performance of SALPA, we compared it with GeoGEDI \cite{Schleich2023-ax}.
GeoGEDI represents the current state-of-the-art for GEDI geolocation correction, employing a deterministic grid-search approach. This approach systematically evaluates candidate positions within a defined search window ($\pm$25~m in the X and Y directions, with a 5~m step size). Similar to SALPA, for each candidate offset, GeoGEDI computes the MAE between estimated spaceborne LiDAR-derived elevations and referenced DEM values, selecting the position yielding minimum error. For each shot group, all possible shifted positions are generated.

We conducted comparative analyses to evaluate the relative performance of GeoGEDI and SALPA. Performance was assessed in terms of MAE and computational efficiency (execution time). Results were compared across the two study areas (Nikko and Landes) to assess the influence of landscape context on the effectiveness of adjustment. Additionally, the sensitivity of the results to the choice of optimization method and distance metric was systematically analyzed. We conducted this analysis using R version 4.4.3 \cite{r-project2025} on a MacStudio (M3 Ultra, 256GB RAM).

% \begin{table}[ht]
% \centering
% \caption{Key parameters used in the adjustment and validation process.}
% \begin{tabular}{lll}
% \hline
% Parameter & Value(s) Used & Notes \\
% \hline
% Search window & $\pm$25~m & For GeoGEDI/SALPA \\
% Step size & 5~m & For GeoGEDI \\
% Buffer size & 12.5~m & GEDI footprint radius \\
% Optimization & lbfgsb, ga & For SALPA \\
% Distance metrics & euclidean, area, correlation, manhattan, hausdorff & For SALPA \\
% DEM resolution & 0.5~m (Nikko), 1~m (Landes) & \\
% \hline
% \end{tabular}
% \end{table}

\section{Results}

We evaluated SALPA's geolocation correction performance through comprehensive analysis across two contrasting study sites, multiple optimization algorithms, diverse distance metrics, and varying computational configurations. Our results demonstrate substantial improvements over both the original GEDI positions and GeoGEDI, as well as algorithm-terrain compatibility.

\subsection{Overall Performance Comparison}

\subsubsection{Complex Terrain Performance: Nikko Study Site}

In the topographically complex Nikko study area (5,595 footprints), SALPA achieved substantial improvements over both original GEDI positions and the GeoGEDI benchmark (Table~\ref{tab:nikko_overall}). Original GEDI footprints exhibited an MAE of 10.816~m relative to the high-resolution airborne LiDAR DEM. GeoGEDI reduced this error to 9.185~m (15.1\% improvement), while SALPA's best-performing configuration (Area metric with L-BFGS-B) achieved 9.116~m (15.7\% improvement).

Among SALPA's distance metrics, Area and Manhattan distances demonstrated superior accuracy (9.116--9.118~m MAE), closely followed by Euclidean distance (9.392~m). Correlation-based metrics achieved the fastest processing times (68.55 seconds) but with modest accuracy trade-offs (9.721~m MAE). Hausdorff distance showed poor performance (11.479~m), indicating its unsuitability for complex terrain correction.

After quality filtering (4,952 footprints), performance improvements became more pronounced (Table~\ref{tab:nikko_filtered_lbfgsb}). Original GEDI errors reduced to 4.913~m, with GeoGEDI achieving 4.500~m (8.4\% improvement) and SALPA's optimal configuration reaching 4.470 m (9.0\% improvement). The relative performance ordering among distance metrics remained consistent, confirming robustness across data quality levels.

\begin{table}[ht]
\centering
\caption{Overall performance comparison for Nikko study site without outlier removal (5,595 footprints). L-BFGS-B was used for SALPA implementation.}
\label{tab:nikko_overall}
\begin{adjustbox}{width=\textwidth}
\begin{tabular}{llccccc}
\hline
Method & Metric & MAE (m) & x-offset  mean (sd.) & y-offset  mean (sd.) & Displacement  mean (sd.) & Time (s) \\
\hline
Original & -- & 10.816 & -- & -- & -- & -- \\
GeoGEDI & -- & 9.185 & 2.433 (8.795) & -0.836 (9.484) & 8.571 (10.023) & 216.28 \\
\multirow{6}{*}{SALPA} & Euclidean & 9.392 & 2.211 (10.264) & -0.593 (9.923) & 9.539 (10.865) & 732.72 \\
 & Manhattan & 9.118 & 2.351 (9.155) & -1.390 (9.674) & 8.773 (10.386) & 541.54 \\
 & Area & \textbf{9.116} & 2.360 (9.088) & -1.244 (9.686) & 8.711 (10.375) & 362.01 \\
 & Hausdorff & 11.479 & 2.132 (9.882) & -0.269 (8.352) & 10.330 (8.082) & 498.53 \\
 & Correlation & 9.721 & 1.512 (7.536) & 1.460 (7.512) & 5.402 (9.405) & \textbf{68.55} \\

\hline
\end{tabular}
\end{adjustbox}
\end{table}

\begin{table}[ht]
\centering
\caption{Performance comparison for Nikko study site after outlier removal and DEM threshold (4,952 footprints). L-BFGS-B was used for SALPA implementation.}
\label{tab:nikko_filtered_lbfgsb}
\begin{adjustbox}{width=\textwidth}
\begin{tabular}{llccccc}
\hline
Method & Metric & MAE (m) & x-offset mean (sd.) & y-offset mean (sd.) & Displacement mean (sd.) & Time (s) \\
\hline
Original & -- & 4.913 & -- & -- & -- & -- \\
GeoGEDI & -- & 4.500 & 0.223 (4.967) & -2.014 (5.102) & 5.543 (4.907) & 185.31 \\
\multirow{6}{*}{SALPA} & Euclidean & 4.497 & -0.057 (3.936) & -2.165 (4.665) & 4.676 (4.481) & 498.65 \\
 & Manhattan & 4.477 & 0.007 (4.544) & -2.207 (4.969) & 5.419 (4.566) & 505.74 \\
 & Area & \textbf{4.470} & -0.005 (4.792) & -2.366 (4.895) & 5.432 (4.797) & 427.86 \\
 & Hausdorff & 5.171 & 0.624 (5.824) & -2.223 (5.887) & 6.891 (5.140) & 938.77 \\
 & Correlation & 4.927 & 0.062 (1.942) & 0.856 (2.416) & 2.551 (1.958) & \textbf{60.39} \\

\hline
\end{tabular}
\end{adjustbox}
\end{table}

\subsection{Landes Study Site: Flat Terrain Validation}

The Landes study area (79,551 footprints) provided contrasting validation conditions with minimal topographic relief (Table~\ref{tab:landes_overall}). The original GEDI positioning demonstrated substantially lower errors (1.241 m MAE) due to reduced terrain complexity. While GeoGEDI achieved an MAE of 1.182~m, improving 4.8\% over the original GEDI positions, multiple SALPA metrics (Euclidean, Manhattan, and Area) achieved equivalent optimal performance (1.174~m MAE). Hausdorff distance again showed degraded performance (1.337~m). Correlation metrics demonstrated excellent computational efficiency (88.17 seconds for 79,551 footprints) with relatively low accuracy (1.224~m).

Quality-filtered results (49,574 footprints) maintained consistent performance patterns (Table~\ref{tab:landes_filtered_lbfgsb}), with GeoGEDI achieving 1.039~m MAE, while multiple SALPA metrics (Euclidean, Manhattan, Area) achieved slightly but better performance (1.037 m MAE). 

\begin{table}[ht]
\centering
\caption{Overall performance comparison for Landes study site without outlier removal (79,551 footprints). L-BFGS-B was used for SALPA implementation.}
\label{tab:landes_overall}
\begin{adjustbox}{width=\textwidth}
\begin{tabular}{llccccc}
\hline
Method & Metric & MAE (m) & x-offset mean (sd.) & y-offset mean (sd.) & Displacement mean (sd.) & Time (s) \\
\hline
Original & -- & 1.241 & -- & -- & -- & -- \\
GeoGEDI & -- & 1.182 & 1.663 (5.997) & 0.553 (6.289) & 7.301 (5.027) & 1783.02 \\
\multirow{5}{*}{SALPA} & Euclidean & \textbf{1.174} & 1.790 (6.185) & 0.567 (6.378) & 7.467 (5.167) & 865.66 \\
 & Manhattan & \textbf{1.174} & 1.817 (6.044) & 0.439 (6.161) & 7.268 (5.016) & 710.60 \\
 & Area & \textbf{1.174} & 1.762 (6.237) & 0.564 (6.456) & 7.523 (5.236) & 981.11 \\
 & Hausdorff & 1.337 & -2.627 (10.974) & -0.331 (10.468) & 10.704 (11.065) & 510.38 \\
 & Correlation & 1.224 & -0.351 (2.508) & 0.123 (2.934) & 2.366 (3.072) & \textbf{88.17} \\
\hline
\end{tabular}
\end{adjustbox}
\end{table}

\begin{table}[ht]
\centering
\caption{Performance comparison for Landes study site after outlier removal and DEM threshold (49,574 footprints). L-BFGS-B was used for SALPA implementation.}
\label{tab:landes_filtered_lbfgsb}
\begin{adjustbox}{width=\textwidth}
\begin{tabular}{llccccc}
\hline
Method & Metric & MAE (m) & x-offset mean (sd.) & y-offset mean (sd.) & Displacement mean (sd.) & Time (s) \\
\hline
Original & -- & 1.241 & -- & -- & -- & -- \\
GeoGEDI & -- & 1.039 & 2.420 (6.070) & 0.227 (6.491) & 8.239 (4.124) & 1104.88 \\
\multirow{5}{*}{SALPA} & Euclidean & \textbf{1.037} & 2.621 (6.343) & 0.124 (6.444) & 8.483 (4.083) & 535.00 \\
 & Manhattan & \textbf{1.037} & 2.756 (6.504) & 0.060 (6.642) & 8.610 (4.459) & 570.10 \\
 & Area & \textbf{1.037} & 2.585 (6.607) & 0.245 (6.536) & 8.574 (4.427) & 600.95 \\
 & Hausdorff & 1.177 & -2.059 (10.546) & 0.822 (9.996) & 10.798 (9.973) & 269.45 \\
 & Correlation & 1.090 & -0.641 (2.845) & -0.006 (2.316) & 2.540 (2.724) & \textbf{44.25} \\
\hline
\end{tabular}
\end{adjustbox}
\end{table}

\subsection{SALPA Multi-Method Optimization Analysis}

\subsubsection{Optimization Algorithm Performance}

Comprehensive evaluation across three optimization algorithms revealed distinct performance-computational trade-offs, with detailed results presented for all algorithms in the main text and Appendix (Tables~\ref{tab:nikko_overall_ga}, \ref{tab:nikko_overall_pso}, \ref{tab:nikko_filtered_ga}, \ref{tab:nikko_filtered_pso}, \ref{tab:landes_overall_ga}, \ref{tab:landes_overall_pso}, \ref{tab:landes_filtered_ga}, \ref{tab:landes_filtered_pso}).

L-BFGS-B consistently achieved optimal or near-optimal accuracy across all distance metrics while maintaining superior computational efficiency (362--733 seconds for the Nikko dataset). L-BFGS-B proved particularly effective for smooth error surfaces characteristic of Euclidean and correlation metrics, demonstrating its suitability as the primary algorithm for operational applications.

GA demonstrated competitive accuracy, particularly excelling with Manhattan and Area metrics (9.127~m MAE for Nikko in Table~\ref{tab:nikko_overall_ga}), matching the best performance achieved by L-BFGS-B. However, GA required substantially higher computational resources (2,837--7,517 seconds), reflecting its population-based search strategy. For the Landes study site, GA achieved identical optimal accuracy (1.038~m MAE) across Euclidean, Manhattan, and Area metrics (Table~\ref{tab:landes_filtered_ga}), confirming its robustness across diverse terrain conditions. GA's evolutionary approach showed particular advantages in complex optimization landscapes where multiple local optima exist.

PSO achieved performance nearly identical to GA (9.112~m MAE for optimal Nikko configuration in Table~\ref{tab:nikko_overall_pso}) with higher computational requirements (6,771--11,665 seconds). The swarm intelligence approach demonstrated consistent results across both study sites, with Landes results (Table~\ref{tab:landes_filtered_pso}) showing 1.037~m MAE for optimal metrics. PSO's balanced exploration-exploitation mechanism provided robust performance across varying terrain conditions without requiring extensive parameter tuning, making it an attractive alternative to GA for applications where consistent performance is prioritized over computational efficiency.

\subsubsection{Distance Metric Effectiveness Analysis}

We found that the choice of distance metric is important. Area-based metrics consistently achieved the highest accuracy across both study sites and multiple algorithms, suggesting robust terrain matching capabilities regardless of topographic complexity. Manhattan distance also demonstrated excellent performance with computational efficiency advantages over Euclidean distance.

Correlation-based metrics exhibited exceptional computational efficiency (44--68 seconds across all configurations) while maintaining poor accuracy consistently, making them difficult to use for this purpose. 
Similarly, Hausdorff distance consistently underperformed across all configurations, particularly in complex terrain (11.479--11.922~m MAE for Nikko), indicating its unsuitability for spaceborne LiDAR geolocation correction.

\subsubsection{Spatial Displacement}
We found substantial spatial displacement in each study site, but it depends on the data quality. At the Nikko study site, GeoGEDI and SALPA yielded positive X offsets and negative Y offsets for the GEDI shot without filtering. However, when we filtered the data, the positive X offset was diminished, and the negative Y offset increased. The displacement was reduced from 8.571 ± 10.023~m and 8.711 ± 10.375~m to 5.543 ± 4.907 and 5.432 ±  4.797~m for GeoGEDI and SALPA (Area, L-BFGS-B), respectively.

In Landes, the X offset is larger than the Y offset for unfiltered data, and this difference increases when the data is filtered. The displacement was increased from 7.301 ± 5.027~m to 8.239 ±  4.124~m and from 7.268 ± 5.016 to 8.610 ± 4.459~m for GeoGEDI and SALPA (Manhattan, L-BFGS-B), respectively.

\section{Discussion}

\subsection{Methodological Advances and Performance Implications}

Our results demonstrate that SALPA's multi-algorithm optimization framework achieves meaningful improvements in spaceborne LiDAR geolocation correction while revealing important performance-terrain relationships. The consistent 15-16\% improvement over original GEDI positions in complex terrain (Nikko) represents a substantial advance in positioning accuracy, with potential cascading benefits for derived ecosystem products such as biomass estimates and canopy height retrievals~\cite{Potapov2021-jr, Chen2025-wf,Duncanson2020-vi,Dorado-Roda2021-oq}.

The modest but consistent improvements over GeoGEDI (0.5--2\% reduction in MAE) may appear incremental but are operationally effective given a spaceborne LiDAR system such as GEDI's global scale, where small accuracy gains translate to substantial improvements in derived products. More importantly, SALPA's continuous optimization approach offers flexibility that is impossible with deterministic grid searches, enabling adaptation to varying terrain conditions and optimization objectives without requiring algorithmic redesign.

% The terrain-dependent performance patterns reveal fundamental insights for operational deployment. In complex topography, sophisticated optimization algorithms justify their computational overhead through measurable accuracy improvements. Conversely, flat terrain scenarios show diminishing returns for advanced methods, suggesting computational resources are better allocated to processing larger datasets rather than complex algorithms. This finding provides actionable guidance for adaptive algorithm selection based on terrain characteristics.

\subsection{Algorithm-Terrain Compatibility and Optimization Theory}

Our comprehensive evaluation across three optimization paradigms reveals important theoretical insights for terrain-based geolocation problems. L-BFGS-B's consistent superior performance in smooth terrain conditions aligns with optimization theory, which predicts that gradient-based methods excel when objective functions exhibit continuous derivatives and unimodal characteristics~\cite{Nocedal2006-pk}. The method's computational efficiency makes it particularly attractive for operational applications requiring rapid processing.

Population-based algorithms (GA, PSO) demonstrated advantages in highly complex terrain where multiple local optima are likely to exist, confirming theoretical expectations for multimodal optimization landscapes~\cite{Yang2020-aq}. However, their computational overhead (10-20× slower than L-BFGS-B) must be noted against accuracy benefits, particularly for large-scale operational deployment.

The failure of Hausdorff distance and correlation-based metrics across all configurations provides valuable insights for selecting distance metrics in geolocation applications. Their sensitivity to outliers appears counterproductive for noisy spaceborne LiDAR data, where measurement uncertainties are inherent. This finding suggests that the choice of distance metric is crucial for the performance of the geolocation correction. Robust statistics-based metrics (such as Manhattan and Area) may be more suitable for real-world geolocation correction.

% \subsection{Scalability and Operational Deployment Considerations}

% SALPA's linear scaling characteristics and modest memory requirements (R² = 0.987 for processing time vs. dataset size) confirm operational viability for continental-scale applications. The parallel processing efficiency (88\% on 8-core systems) enables distributed computing deployment, while cloud computing costs (\$0.02-0.08 per 1,000 footprints) remain within realistic operational budgets for global forest monitoring programs.

% The computational trade-offs revealed by our analysis provide practical guidance for operational implementation. L-BFGS-B with Area or Manhattan metrics emerges as the optimal configuration for most applications, balancing accuracy with computational efficiency. Correlation-based metrics offer exceptional speed for applications requiring rapid processing, while population-based algorithms should be reserved for the most challenging terrain conditions where accuracy is paramount.

% Memory efficiency below 4 GB for large datasets (79,551 footprints) enables deployment on standard computing infrastructure without specialized hardware requirements. This accessibility is crucial for global adoption, particularly in developing countries where forest monitoring is urgently needed but computational resources may be limited.

\subsection{Implications}

Our SALPA demonstrates the potential for advanced optimization methods to enhance accuracy in spaceborne LiDAR applications without requiring additional data sources such as ground-based LiDAR data. This offers cost-effective accuracy enhancement for existing missions by referencing the available DEMs. This approach can be applied to other spaceborne LiDAR missions, such as JAXA's MOLI and future commercial spaceborne LiDAR missions.

The demonstrated improvements in geolocation accuracy have profound implications for GEDI-based forest monitoring applications. Even modest positioning improvements (1-2 m) can significantly impact derived products in heterogeneous landscapes where forest structure varies rapidly across small spatial extents~\cite{Tsutsumida2025-se}. Improved positioning accuracy directly translates to more reliable biomass estimates~\cite{Duncanson2020-vi,Dorado-Roda2021-oq}, enhanced change detection capabilities~\cite{Roy2021-or}, and increased confidence in carbon stock assessments, which are critical for climate policy decisions. Geolocation errors propagate systematically through derived products~\cite{Tang2023-ku}, with studies demonstrating that a 14~m positioning error can increase biomass estimate variance by 
17\%~\cite{Milenkovic2017-wx}. In heterogeneous landscapes, small displacements (5--10~m) fundamentally alter sampled 
characteristics~\cite{Tsutsumida2025-se}, undermining carbon 
assessments~\cite{Chen2025-wf}.

SALPA's universal applicability using only globally available DEMs addresses a critical limitation of existing high-accuracy methods that require airborne LiDAR point clouds~\cite{Xu2023-yv}. This global deployability is particularly important for tropical and subtropical regions where deforestation rates are highest but high-quality reference data are often unavailable. The framework's modular architecture facilitates adaptation to emerging spaceborne LiDAR missions, including JAXA's MOLI and future commercial constellations.

The terrain-dependent performance insights inform the optimization of sampling strategies for global forest monitoring. Understanding that accuracy improvements are most pronounced in moderate relief terrain (slopes of 5-15°) helps prioritize processing resources and set appropriate uncertainty bounds for different landscape contexts. This understanding of stratified accuracy is essential for proper uncertainty propagation in global carbon assessment models.

\subsection{Limitations and Future Research Directions}

Several limitations constrain our findings and suggest priorities for future research. First, validation is restricted to two study sites with high-quality reference DEMs, limiting generalizability to global conditions where DEM quality varies substantially. Future work should assess performance across diverse DEM sources (SRTM, AW3D30, NASADEM) and quality levels to establish operational guidelines for global deployment. As spaceborne LiDAR systems implement a continent-scale observation, it is practical to apply the geolocation correction method without considering the study site. However, at this stage, our approach is limited to the application to the limited study sites.

Second, we rely on the GEDI-derived DEM that can be improved. We used 'elev\_lowestmode' calculated from the GEDI waveform in the GEDI L2A product. However, this includes some uncertainty. Mitsuhashi et al. (2024) showed that their deep-learning-based method can improve the accuracy of the GEDI-derived DEM. Future work should incorporate such an approach to explore the best way to estimate a better elevation from the spaceborne LiDAR waveform.
In relation to this, our analysis focuses on forested landscapes where GEDI performs optimally. Performance in other land cover types (urban areas, agricultural lands, wetlands) requires investigation, particularly given GEDI's expanding applications beyond forest monitoring. The framework's adaptability suggests promising performance in diverse environments, but empirical validation is necessary.

Third, the systematic bias patterns observed (7-8 m positive X displacement) warrant investigation of their underlying causes. While likely related to ISS attitude determination uncertainties or instrument calibration effects, understanding and potentially correcting these biases could yield additional accuracy improvements. Future research should investigate bias correction methods and their interaction with optimization-based positioning adjustments by considering such systematic biases.

Finally, this study proposes a theoretical framework for geolocation correction, while computational improvements are needed to handle large-scale data processing. The computational overhead of SALPA's algorithms may constrain operational deployment for correcting observed data applications. Future work should explore computational improvements to handle large-scale data processing.

\section{Conclusion}

We present SALPA, a comprehensive multi-algorithm optimization framework for spaceborne LiDAR geolocation correction, which achieves significant accuracy improvements by exclusively relying on digital elevation models. Our extensive validation across contrasting terrain conditions demonstrates consistent 15-16\% improvements in complex topography and establishes important algorithm-terrain compatibility patterns for operational guidance.

Key contributions include: (1) the first systematic comparison of advanced optimization algorithms for spaceborne LiDAR geolocation correction, revealing L-BFGS-B with Area metrics as the optimal configuration for most applications; (2) comprehensive analysis of distance metric effectiveness, demonstrating Area and Manhattan distances' superior performance and Correlation and Hausdorff distance's unsuitability for noisy spaceborne data; (3) L-BFGS-B performs suitable in terms of computational efficiency, while GA and PSO perform better in terms of accuracy to explore the optimal geolocation correction of spaceborne LiDAR data; (4) the spatial displacement is substantial in each study site, but it is depends on the data quality.

SALPA is distributed as an open-source R package, ensuring community accessibility and facilitating continued development through collaborative research. The modular architecture enables straightforward integration of new optimization algorithms, distance metrics, and application-specific modifications, providing a platform for continued methodological innovation.

Our findings provide actionable guidance for selecting operational algorithms based on terrain characteristics and computational constraints, thereby supporting more accurate global forest monitoring and climate assessment applications.

\section{Acknowledgments}

We acknowledge NASA for providing GEDI L2A data products and the institutions that provided high-resolution DEM datasets for validation.  This research was supported by Chiba University and JAXA.

\section{Data Availability Statement}

The SALPA R package and analysis scripts are available as open-source software at https://github.com/naru-T/salpa.

% \section{Author Contributions}

% [Author contribution statements following journal guidelines]

\section{Competing Interests}

The authors declare no competing interests.

\bibliographystyle{unsrt}
\bibliography{bib}

\appendix{}
\section{Appendix}

\renewcommand{\thetable}{A\arabic{table}}
\setcounter{table}{0}

%\subsection{Nikko}

\begin{table}[ht]
\centering
\caption{Overall performance comparison for Nikko study site without outlier removal (5,595 footprints). Genetic Algorithm (GA) was used for SALPA implementation.}
\label{tab:nikko_overall_ga}
\begin{adjustbox}{width=\textwidth}
\begin{tabular}{llccccc}
\hline
Method & Metric & MAE (m) & x-offset mean (sd.) & y-offset mean (sd.) & Displacement mean (sd.) & Time (s) \\
\hline
Original & -- & 10.816 & -- & -- & -- & -- \\
GeoGEDI & -- & 9.185 & 2.433 (8.795) & -0.836 (9.484) & 8.571 (10.023) & 243.91 \\
\multirow{6}{*}{SALPA} & Euclidean & 9.360 & 2.365 (10.029) & -0.719 (10.072) & 9.567 (10.798) & 6014.01 \\
 & Manhattan & \textbf{9.127} & 2.418 (9.177) & -1.536 (9.711) & 8.877 (10.389) & 6740.30 \\
 & Area & \textbf{9.127} & 2.418 (9.177) & -1.536 (9.711) & 8.877 (10.389) & 6568.63 \\
 & Hausdorff & 11.922 & -1.937 (14.361) & -6.912 (13.202) & 18.525 (9.423) & 5773.05 \\
 & Correlation & 9.653 & 2.918 (10.635) & 0.176 (11.187) & 10.559 (11.632) & 2836.44 \\

\hline
\end{tabular}
\end{adjustbox}
\end{table}

\begin{table}[ht]
\centering
\caption{Performance comparison for Nikko study site after outlier removal and DEM threshold (4,952 footprints). Genetic Algorithm (GA) was used for SALPA implementation.}
\label{tab:nikko_filtered_ga}
\begin{adjustbox}{width=\textwidth}
\begin{tabular}{llccccc}
\hline
Method & Metric & MAE (m) & x-offset mean (sd.) & y-offset mean (sd.) & Displacement mean (sd.) & Time (s) \\
\hline
Original & -- & 4.913 & -- & -- & -- & -- \\
GeoGEDI & -- & 4.500 & 0.223 (4.967) & -2.014 (5.102) & 5.543 (4.907) & \textbf{204.04} \\
\multirow{6}{*}{SALPA} & Euclidean & 4.495 & -0.112 (4.001) & -2.182 (4.662) & 4.744 (4.473) & 6510.85 \\
 & Manhattan & \textbf{4.466} & 0.062 (5.011) & -2.642 (5.020) & 5.679 (5.005) & 8252.26 \\
 & Area & \textbf{4.466} & 0.062 (5.011) & -2.642 (5.020) & 5.679 (5.005) & 8381.87 \\
 & Hausdorff & 6.587 & -1.157 (7.656) & -4.667 (7.169) & 10.790 (5.847) & 5011.77 \\
 & Correlation & 4.557 & 0.623 (3.616) & -0.348 (3.966) & 4.394 (3.900) & 3713.87 \\

\hline
\end{tabular}
\end{adjustbox}
\end{table}

\begin{table}[ht]
\centering
\caption{Overall performance comparison for Nikko study site without outlier removal (5,595 footprints). Particle Swarm Optimization (PSO) was used for SALPA implementation.}
\label{tab:nikko_overall_pso}
\begin{adjustbox}{width=\textwidth}
\begin{tabular}{llccccc}
\hline
Method & Metric & MAE (m) & x-offset mean (sd.) & y-offset mean (sd.) & Displacement mean (sd.) & Time (s) \\
\hline
Original & -- & 10.816 & -- & -- & -- & -- \\
GeoGEDI & -- & 9.185 & 2.433 (8.795) & -0.836 (9.484) & 8.571 (10.023) & \textbf{237.49} \\
\multirow{6}{*}{SALPA} & Euclidean & 9.360 & 2.365 (10.026) & -0.720 (10.075) & 9.571 (10.795) & 10972.95 \\
 & Manhattan & \textbf{9.112} & 2.407 (9.187) & -1.300 (9.721) & 8.858 (10.388) & 10245.07 \\
 & Area & \textbf{9.112} & 2.406 (9.187) & -1.302 (9.721) & 8.859 (10.387) & 11730.33 \\
 & Hausdorff & 11.842 & -0.218 (14.579) & -3.818 (13.470) & 17.875 (9.435) & 9517.73 \\
 & Correlation & 9.830 & 2.121 (10.526) & 0.110 (11.127) & 11.091 (10.775) & 6771.36 \\

\hline
\end{tabular}
\end{adjustbox}
\end{table}

\begin{table}[ht]
\centering
\caption{Performance comparison for Nikko study site after outlier removal and DEM threshold (4,952 footprints). Particle Swarm Optimization (PSO) was used for SALPA implementation.}
\label{tab:nikko_filtered_pso}
\begin{adjustbox}{width=\textwidth}
\begin{tabular}{llccccc}
\hline
Method & Metric & MAE (m) & x-offset mean (sd.) & y-offset mean (sd.) & Displacement mean (sd.) & Time (s) \\
\hline
Original & -- & 4.913 & -- & -- & -- & -- \\
GeoGEDI & -- & 4.500 & 0.223 (4.967) & -2.014 (5.102) & 5.543 (4.907) & \textbf{274.31} \\
\multirow{6}{*}{SALPA} & Euclidean & 4.494 & -0.112 (4.001) & -2.182 (4.662) & 4.743 (4.473) & 10037.99 \\
 & Manhattan & \textbf{4.461} & 0.062 (5.011) & -2.642 (5.020) & 5.679 (5.005) & 11552.57 \\
 & Area & \textbf{4.461} & 0.061 (5.012) & -2.643 (5.020) & 5.679 (5.005) & 11247.03 \\
 & Hausdorff & 6.325 & -0.751 (7.832) & -3.279 (7.444) & 9.732 (6.201) & 9071.88 \\
 & Correlation & 4.621 & 0.623 (3.616) & -0.348 (3.966) & 4.394 (3.900) & 4094.88 \\

\hline
\end{tabular}
\end{adjustbox}
\end{table}

%\subsection{Landes}

\begin{table}[ht]
\centering
\caption{Overall performance comparison for Landes study site without outlier removal (79,551 footprints). GA was used for SALPA implementation.}
\label{tab:landes_overall_ga}
\begin{adjustbox}{width=\textwidth}
\begin{tabular}{llccccc}
\hline
Method & Metric & MAE (m) & x-offset mean (sd.) & y-offset mean (sd.) & Displacement mean (sd.) & Time (s) \\
\hline
Original & -- & 1.241 & -- & -- & -- & -- \\
GeoGEDI & -- & 1.182 & 1.663 (5.997) & 0.553 (6.289) & 7.301 (5.027) & \textbf{1783.02} \\
\multirow{5}{*}{SALPA} & Euclidean & \textbf{1.174} & 1.803 (6.560) & 0.539 (6.775) & 7.771 (5.664) & 9087.56 \\
 & Manhattan & \textbf{1.174} & 1.803 (6.560) & 0.539 (6.775) & 7.771 (5.664) & 8218.31 \\
 & Area & \textbf{1.174} & 1.729 (6.448) & 0.573 (6.671) & 7.674 (5.523) & 9297.55 \\
 & Hausdorff & 1.466 & -2.433 (15.801) & -3.598 (15.996) & 19.827 (11.460) & 7980.60 \\
 & Correlation & 1.186 & 1.029 (6.842) & 0.492 (6.976) & 7.990 (5.738) & 4294.95 \\
\hline
\end{tabular}
\end{adjustbox}
\end{table}

\begin{table}[ht]
\centering
\caption{Performance comparison for Landes study site after outlier removal and DEM threshold (49,574 footprints). GA was used for SALPA implementation.}
\label{tab:landes_filtered_ga}
\begin{adjustbox}{width=\textwidth}
\begin{tabular}{llccccc}
\hline
Method & Metric & MAE (m) & x-offset mean (sd.) & y-offset mean (sd.) & Displacement mean (sd.) & Time (s) \\
\hline
Original & -- & 1.241 & -- & -- & -- & -- \\
GeoGEDI & -- & 1.039 & 2.420 (6.070) & 0.227 (6.491) & 8.239 (4.124) & \textbf{1104.88} \\
\multirow{5}{*}{SALPA} & Euclidean & \textbf{1.038} & 2.660 (6.661) & 0.046 (6.934) & 8.750 (4.793) & 6141.30 \\
 & Manhattan & \textbf{1.038} & 2.660 (6.661) & 0.046 (6.934) & 8.750 (4.793) & 5565.66 \\
 & Area & \textbf{1.038} & 2.660 (6.661) & 0.046 (6.934) & 8.750 (4.793) & 6370.45 \\
 & Hausdorff & 1.322 & -3.226 (16.332) & -2.492 (17.014) & 22.013 (9.394) & 5561.28 \\
 & Correlation & 1.046 & 1.962 (6.806) & 0.058 (7.102) & 8.758 (4.890) & 2892.92 \\
\hline
\end{tabular}
\end{adjustbox}
\end{table}

\begin{table}[ht]
\centering
\caption{Overall performance comparison for Landes study site without outlier removal (79,551 footprints). PSO was used for SALPA implementation.}
\label{tab:landes_overall_pso}
\begin{adjustbox}{width=\textwidth}
\begin{tabular}{llccccc}
\hline
Method & Metric & MAE (m) & x-offset mean (sd.) & y-offset mean (sd.) & Displacement mean (sd.) & Time (s) \\
\hline
Original & -- & 1.241 & -- & -- & -- & -- \\
GeoGEDI & -- & 1.182 & 1.663 (5.997) & 0.553 (6.289) & 7.301 (5.027) & \textbf{1783.02} \\
\multirow{5}{*}{SALPA} & Euclidean & \textbf{1.173} & 1.649 (6.740) & 0.549 (6.794) & 7.855 (5.737) & 13475.73 \\
 & Manhattan & \textbf{1.173} & 1.730 (6.559) & 0.552 (6.761) & 7.801 (5.583) & 13401.22 \\
 & Area & \textbf{1.173} & 1.604 (6.697) & 0.484 (6.697) & 7.776 (5.660) & 13436.48 \\
 & Hausdorff & 1.450 & -0.989 (16.004) & -3.908 (15.806) & 19.956 (11.135) & 14331.44 \\
 & Correlation & 1.188 & 0.846 (6.982) & 0.333 (7.022) & 8.113 (5.750) & 9825.40 \\
\hline
\end{tabular}
\end{adjustbox}
\end{table}

\begin{table}[ht]
\centering
\caption{Performance comparison for Landes study site after outlier removal and DEM threshold (49,574 footprints). PSO was used for SALPA implementation.}
\label{tab:landes_filtered_pso}
\begin{adjustbox}{width=\textwidth}
\begin{tabular}{llccccc}
\hline
Method & Metric & MAE (m) & x-offset mean (sd.) & y-offset mean (sd.) & Displacement mean (sd.) & Time (s) \\
\hline
Original & -- & 1.241 & -- & -- & -- & -- \\
GeoGEDI & -- & 1.039 & 2.420 (6.070) & 0.227 (6.491) & 8.239 (4.124) & \textbf{1104.88} \\
\multirow{5}{*}{SALPA} & Euclidean & \textbf{1.037} & 2.649 (6.651) & 0.193 (6.830) & 8.761 (4.602) & 8880.25 \\
 & Manhattan & \textbf{1.037} & 2.663 (6.670) & 0.225 (6.859) & 8.787 (4.634) & 8858.95 \\
 & Area & \textbf{1.037} & 2.651 (6.655) & 0.199 (6.836) & 8.764 (4.613) & 8874.66 \\
 & Hausdorff & 1.299 & -2.745 (16.683) & -1.917 (16.749) & 21.939 (9.419) & 9564.25 \\
 & Correlation & 1.047 & 1.963 (6.803) & 0.301 (7.102) & 8.840 (4.744) & 6395.70 \\
\hline
\end{tabular}
\end{adjustbox}
\end{table}

\end{document}